\documentclass[showpacs,aps,floatfix,preprint]{revtex4-1}
\usepackage{amssymb}
\usepackage{amsmath}
\usepackage{graphicx}
\usepackage{dcolumn}
\usepackage{amsfonts}
\usepackage{bm}
\newcommand{\bra}[1]{\left\langle #1\right|}
\newcommand{\ket}[1]{\left| #1\right\rangle}
\begin{document}

\title{Nonlocal dispersive optical model ingredients for ${}^{40}$\textrm{Ca}}
\author{M. H. Mahzoon$^{1}$, R. J. Charity$^{2}$, W. H. Dickhoff$^{1}$, H. Dussan$^1$, S. J. Waldecker$^3$}
\affiliation{Department of Physics$^{1}$ and Chemistry$^{2}$, Washington University, St.
Louis, Missouri 63130}
\affiliation{Department of Physics$^3$, University of Tennessee, Chattanooga, TN 37403}
\date{\today}

\begin{abstract}
A comprehensive description of all single-particle properties associated with the nucleus ${}^{40}$Ca has been generated by employing a nonlocal dispersive optical potential capable of simultaneously reproducing all relevant data above and below the Fermi energy.
We gather all relevant functional forms and the numerical values of the parameters in this contribution.
\end{abstract}

\maketitle

\section{Introduction}
\label{Sec:intro}
Supplementary information is gathered here to provide the equations and parameters associated with a short paper submitted for publication~\cite{Hossein13b}.
We employ a new version of the dispersive optical model (DOM) first developed by Mahaux and Sartor~\cite{Mahaux91}.
The present implementation is geared towards treating the potential as a practical representation of the nucleon self-energy of ${}^{40}$Ca.
The aim is to represent all available data related to the single-particle propagator~\cite{Dickhoff08}.
As in previous versions of the DOM~\cite{Charity06,Charity07,Mueller11}, these data include all available elastic-scattering cross sections up to 200 MeV for both protons and neutrons.
Based on recent theoretical work~\cite{Dickhoff10,Waldecker2011,Dussan11}, we now incorporate a nonlocal representation of both the real Hartree-Fock (HF) as well as the imaginary absorptive potentials.
We have found that these nonlocal potentials are essential to represent data below the Fermi energy that probe properties of the ground state.
Such data cover the location of the main levels of ${}^{40}$Ca both near the Fermi energy as well as for deeply-bound orbits, the charge density, and the spectral function for the removal of high-momentum protons~\cite{Rohe04,Rohe04A}.
A critical and nontrivial constraint to obtain meaningful results is associated with particle number for both protons and neutrons.
We view the construction of the nonlocal DOM potential as a critical 
link between nuclear reactions and nuclear structure as it represents different energy domains of the one underlying nucleon self-energy.
This is accomplished by employing dispersion relations linking the physics above and below the Fermi energy with both sides being influenced by the absorptive potentials on the opposite side.

\section{Formalism}
\label{Sec:form}
The self-energy $\Sigma_{\ell j}$ allows for the solution of the Dyson equation for the nucleon propagator $G_{\ell j}$.  
In the angular-momentum basis
\begin{eqnarray}
\label{eq:dyson}
G_{\ell j}(r,r';E) &=& G^{(0)}_{\ell j}(r,r';E) + \int \!\! d\tilde{r}\ \tilde{r}^2 \!\! \int \!\! d\tilde{r}'\ \tilde{r}'^2 G^{(0)}_{\ell j}(r,\tilde{r};E)
\Sigma_{\ell j}(\tilde{r},\tilde{r}';E) G_{\ell j}(\tilde{r}',r';E) .
\end{eqnarray}
The noninteracting propagators $G^{(0)}_{\ell j}$ only contain kinetic-energy contributions.
The solution of this equation generates $S_{\ell j}(r;E) =   \textrm{Im}\ G_{\ell j}(r,r;E)/\pi$, the hole spectral density, 
for negative continuum energies.
The spectral strength at $E$, for a given $\ell j$, is given by 
\begin{equation}
S_{\ell j}(E) = \int_{0}^\infty dr\ r^2\ S_{\ell j}(r;E) .
\label{eq:specs}
\end{equation}
For discrete energies, one solves the eigenvalue equation for the overlap functions
\begin{equation}
\psi^n_{\ell j}(r) = \bra{\Psi^{A-1}_n}a_{r \ell j} \ket{\Psi^A_0},
\label{eq:overlap}
\end{equation}
for the removal of a nucleon at $r$ with discrete quantum numbers $\ell$ and $j$~\cite{Dickhoff10}.
The removal energy corresponds to
\begin{equation}
\varepsilon^-_n=E^A_0 -E^{A-1}_n
\label{eq:eig}
\end{equation}
with normalization for such a quasi hole state $\alpha_{qh}$ given by
\begin{equation}
S^n_{\ell j} = \left( 1 - \left. \frac{
\partial \Sigma_{\ell j}(\alpha_{qh},
\alpha_{qh}; E)}{\partial E} \right|_{\varepsilon^-_n}\right)^{-1} .
\label{eq:sfac}
\end{equation}
We note that from the solution of the Dyson equation below the Fermi energy, one can generate the one-body density matrix by integrating the non-diagonal imaginary part of the propagator up to the Fermi energy and therefore access expectation values of one-body operators in the ground state including particle number, kinetic energy and charge density~\cite{Dickhoff08}. 
The latter is obtained by folding the point density with the nucleon form factors following the procedure of Ref.~\cite{Brown79}.
For positive scattering energies, it was already realized long ago that the reducible self-energy provides the scattering amplitude for elastic nucleon scattering~\cite{Bell59}.

The nucleon self-energy fulfills the dispersion relation which relates the physics of bound nucleons to those that propagate at positive energy~\cite{Dickhoff08}
\begin{eqnarray} 
\mbox{Re}\ \Sigma_{\ell j}(r,r';E)\! = \! \Sigma^s_{\ell j} (r,r')\! - \! {\cal P} \!\!
\int_{\varepsilon_F^+}^{\infty} \!\! \frac{dE'}{\pi} \frac{\mbox{Im}\ \Sigma_{\ell j}(r,r';E')}{E-E'}  
+{\cal P} \!\!
\int_{-\infty}^{\varepsilon_F^-} \!\! \frac{dE'}{\pi} \frac{\mbox{Im}\ \Sigma_{\ell j}(r,r';E')}{E-E'} .
\label{eq:disprel}
\end{eqnarray}
It contains a static and real correlated HF-like term $\Sigma^s_{\ell j}$ and dynamic parts governed by $\mbox{Im}\Sigma_{\ell j}$ representing the coupling in the $A\pm1$ systems that start and end at the Fermi energies for addition ($\varepsilon_F^+ = E^{A+1}_0-E^A_0$) and removal ($\varepsilon_F^-=E^A_0-E^{A-1}_0$), respectively.
The latter feature is particular to a finite system and allows for discrete quasi particle and hole solutions of the Dyson equation where the imaginary part of the self-energy vanishes.
It is convenient to introduce the average Fermi energy
\begin{equation}
\varepsilon_F = \frac{1}{2} \left[
\varepsilon_F^+  - \varepsilon_F^- \right]
\label{Eq:Fermi}
\end{equation}
and employ the subtracted form of the dispersion relation calculated at this energy~\cite{Mahaux91,Dickhoff10}
\begin{eqnarray} 
\mbox{Re}\ \Sigma_{\ell j}(r,r';E)\! = \!  \Sigma_{\ell j} (r,r';\varepsilon_F)  
- \! {\cal P} \!\!
\int_{\varepsilon_T^+}^{\infty} \!\! \frac{dE'}{\pi} \mbox{Im}\ \Sigma_{\ell j}(r,r';E') \left[ \frac{1}{E-E'}  - \frac{1}{\varepsilon_F -E'} \right]  \nonumber  \\
+{\cal P} \!\!
\int_{-\infty}^{\varepsilon_T^-} \!\! \frac{dE'}{\pi} \mbox{Im}\ \Sigma_{\ell j}(r,r';E') \left[ \frac{1}{E-E'}
-\frac{1}{\varepsilon_F -E'} \right]  ,
 \label{eq:sdisprel} 
\end{eqnarray}
where $\mathcal{P}$ represents the principal value.
The beauty of this representation was recognized by Mahaux and Sartor~\cite{Mahaux86,Mahaux91} since it allows for a link with empirical information both for the real part of the nonlocal self-energy at the Fermi energy (probed by a multitude of HF calculations) as well as through empirical knowledge of the imaginary part of the optical potential also constrained by experimental data.
Consequently Eq.~(\ref{eq:sdisprel}) yields a dynamic contribution to the real part linking \textrm{both} energy domains around the Fermi energy.
Equation~(\ref{eq:sdisprel}) also emphasizes empirical information near the Fermi energy because of the $E'^{-2}$-weighting in the integrands.
The real self-energy at the Fermi energy will be denoted in the following by $\Sigma_{HF}$.

\section{Parametrization of the potentials}
\label{Sec:param}
We now provide a more detailed description of the DOM functionals in order for the resulting potential to yield a realistic description of the single-particle properties below the Fermi energy.
The subtracted dispersion relation shown in Eq.~(\ref{eq:sdisprel}) clarifies that we need a representation of $\Sigma_{HF}$ and the imaginary part of the self-energy $\textrm{Im}\ \Sigma(E)$ at all energies.
We will use a simple Gaussian nonlocality as first suggested in Ref.~\cite{Perey62} in all instances.
Microscopically calculated self-energies exhibit more complicated forms but integrated quantities like volume integrals appear not to require these~\cite{Waldecker2011,Dussan11}.
We restrict the nonlocal contributions to the HF term and the volume and surface contributions to the imaginary part of the potential.
We write the HF self-energy term in the following form
\begin{eqnarray}
\Sigma_{HF}(\bm{r},\bm{r}') = \Sigma^{nl}_{HF}(\bm{r},\bm{r}') + \delta(\bm{r}-\bm{r}') V_C(r) + \delta(\bm{r}-\bm{r}') V^{so}_l(r) ,
\label{eq:HFc}
\end{eqnarray}
with the Coulomb and local spin-orbit contributions.
The nonlocal term is split into a volume and a wine-bottle shape generating contribution
\begin{eqnarray}
\Sigma_{HF}^{nl}\left( \bm{r},\bm{r}' \right)   =   -V_{HF}^{vol}\left( \bm{r},\bm{r}'\right) 
+ V_{HF}^{wb}(\bm{r},\bm{r}') ,
\label{eq:HFn}
\end{eqnarray}
where the volume term is given by
\begin{eqnarray}
V_{HF}^{vol}\left( \bm{r},\bm{r}' \right) = V_{HF}^0
\,f \left ( \tilde{r},r^{HF},a^{HF} \right )  \left [ x_1 H \left( \bm{s};\beta_{vol_1} \right) + (1-x_1) H \left( \bm{s};\beta_{vol_2}\right) \right ] ,
 \label{eq:HFvol} 
\end{eqnarray}
allowing for two different nonlocalities with different weight ($0 \le x_1 \le1$). 
We use the notation $\tilde{r} =(r+r')/2$ and $\bm{s}=\bm{r}-\bm{r}'$.
The wine bottle ($wb$) shape producing Gaussian replaces the surface term of Ref.~\cite{Mueller11}
\begin{equation}
V_{HF}^{wb}(\bm{r},\bm{r}') = V_{wb}^0  \exp{\left(- \tilde{r}^2/\rho_{wb}^2\right)} H \left( \bm{s};\beta_{wb} \right ).
\label{eq:wb}
\end{equation}
This Gaussian centered at the origin also helps to represent overlap functions generated by simple potentials that reproduce corresponding Monte Carlo results~\cite{Brida11}. 
Non-locality is represented by a Gaussian form
\begin{equation}
H \left( \bm{s}; \beta \right) = \exp \left( - \bm{s}^2 / \beta^2 \right)/ (\pi^{3/2} \beta^3)
\end{equation}
first suggested in Ref.~\cite{Perey62}.
As usual we employ Woods-Saxon form factors 
\begin{eqnarray}
f(r,r_{i},a_{i})=\left[1+\exp \left({\frac{r-r_{i}A^{1/3}}{a_{i}}%
}\right)\right]^{-1} .
\label{Eq:WS}
\end{eqnarray}
The Coulomb term is obtained from the calculated charge density and no longer by the potential from a homogeneous sphere as in all previous work (see \textit{e.g.} Ref.~\cite{Mueller11}).
The local spin-orbit interaction is given by
\begin{eqnarray}
V^{so}_{l}(r)= \left( \frac{\hbar}{m_{\pi }c}\right)
^{2} V^{so}
 \frac{1}{r}\frac{d}{dr}f(r,r^{so},a^{so})\; \bm{\ell}\cdot \bm{\sigma},
\label{eq:HFso}
\end{eqnarray}
where $\left( \hbar /m_{\pi }c\right) ^{2}$=2.0~fm$^{2}$ 
as in Ref.~\cite{Mueller11}.

The introduction of nonlocality in the imaginary part of the self-energy is well-founded theoretically both for long-range correlations~\cite{Waldecker2011} as well short-range ones~\cite{Dussan11}. 
Its implied $\ell$-dependence is essential in reproducing the correct particle number for protons and neutrons.
The assumed imaginary component of the potential has the form
\begin{eqnarray}
\textrm{Im}\ \Sigma( \bm{r},\bm{r}',E) =   \textrm{Im}\ \Sigma^{nl}(\bm{r},\bm{r}';E) + \delta(\bm{r}-\bm{r}') \mathcal{W}^{so}_l(r;E) .
\label{Eq:imsig}
\end{eqnarray}
The nonlocal contribution is represented by
\begin{eqnarray}
\textrm{Im}\ \Sigma^{nl}(\bm{r},\bm{r}';E) &=& 
-W^{vol}_{0\pm}(E) f\left(\tilde{r};r^{vol}_{\pm};a^{vol}_{\pm}\right)H \left( \bm{s}; \beta^{\pm}_{vol}\right)  \nonumber \\
&+& 4a^{sur}_{\pm}W^{sur}_{\pm}\left( E\right)H \left( \bm{s}; \beta^{\pm}_{sur}\right) \frac{d}{d \tilde{r} }f(\tilde{r},r^{sur}_{\pm},a^{sur}) .
\label{eq:imnl}
\end{eqnarray}
The phase space of particle levels for $E\gg \varepsilon_F$ is significantly larger
than that of hole levels for $E\ll \varepsilon_F$. Therefore the density of states of 
two-particle-one-hole states and more complicated states for $E\gg \varepsilon_F$ in the self-energy will be larger than the one for two-hole-one-particle states (and more complicated states) at $E\ll \varepsilon_F$. Thus at energies well removed
from $\varepsilon_F$, the form of the imaginary volume potential should not be
symmetric about $\varepsilon_F$ as indicated by the $\pm$ notation~\cite{Dussan11}.
While more symmetric about $\varepsilon_F$, we have allowed a similar option for the surface absorption that is also supported by theoretical work reported in Ref.~\cite{Waldecker2011}.
We include a local spin-orbit contribution with the same form as in Eq.~(\ref{eq:HFso})
\begin{eqnarray}
\mathcal{W}^{so}(r,E)=  \left( \frac{\hbar}{m_{\pi }c}\right)
^{2}W^{so}(E) 
 \frac{1}{r}\frac{d}{dr}f(r,r^{so},a^{so})\; \bm{\ell}\cdot \bm{\sigma},
 \label{eq:IMso}
\end{eqnarray}
using the same geometry parameters as in Eq.~(\ref{eq:HFso}) following Ref.~\cite{Mueller11}.
Allowing for the aforementioned asymmetry around $\varepsilon_F$ the following form was assumed for 
the depth of volume potential~\cite{Mueller11}
\begin{equation}
W^{vol}_{0\pm}(E) =  \Delta W^{\pm}_{NM}(E) +  
\begin{cases}
0 & \text{if } |E-\varepsilon_F| < E^{vol}_{p\pm} \\
A^{vol}_{\pm} \frac{\left(|E-\varepsilon_F|-E^{vol}_{p\pm}\right)^4}
{\left(|E-\varepsilon_F|-E^{vol}_{p\pm}\right)^4 + (B^{vol}_{\pm})^4} & 
 \text{if } |E-\varepsilon_F| > E^{vol}_{p\pm} ,
\end{cases} 
\label{eq:volumeS}
\end{equation}
where $\Delta W^{\pm}_{NM}(E)$ is the energy-asymmetric correction modeled after
nuclear-matter calculations. Apart from this correction, the parametrization
is similar to the Jeukenne and Mahaux form~\cite{Jeukenne83} used in many DOM\
analyses.
The asymmetry above and below $\varepsilon_F$ is essential to accommodate the Jefferson Lab $(e,e'p)$ data at large missing energy.
The energy-asymmetric correction was taken as 
\begin{widetext} 
\begin{equation}
\Delta W^{\pm}_{NM}(E)=
\begin{cases}
\alpha A^{vol}_+ \left[ \sqrt{E}+\frac{\left( \varepsilon_F+E^+_{a}\right) ^{3/2}}{2E}-\frac{3}{2}
\sqrt{\varepsilon_F+E^+_{a}}\right] & \text{for }E-\varepsilon_F>E^+_{a} \\ 
- A^{vol}_- \frac{(\varepsilon_F-E-E^-_{a})^2}{(\varepsilon_F-E-E^-_{a})^2+(E^-_{a})^2} & \text{for }E-\varepsilon_{F}<-E^-_{a} \\ 
0 & \text{otherwise}
\end{cases}
\label{eq:Wnmnl}
\end{equation}
\end{widetext}
which is similar to the form suggested by Mahaux and Sartor~\cite{Mahaux91}.

To describe the energy dependence of surface absorption we employed the form of Ref.~\cite{Charity07}
\begin{eqnarray}
W^{sur}_{\pm}\left( E\right) =\omega _{4}(E,A^{sur}_{\pm},B^{sur}_{\pm s1},0)-\omega
_{2}(E,A^{sur}_{\pm},B_{\pm s2}^{sur},C^{sur}_{\pm}),  \label{eq:paranl} 
\end{eqnarray}
where
\begin{eqnarray}
\omega _{n}(E,A^{sur},B^{sur},C^{sur})=A^{sur}\;\Theta \left(
X\right) \frac{X^{n}}{X^{n}+\left( B^{sur}\right) ^{n}},
\label{eq:omega}
\end{eqnarray}%
and $\Theta \left( X\right) $ is Heaviside's step function and $%
X=\left\vert E-\varepsilon_F\right\vert -C^{sur}$. The functions $\omega _{n}$ are
very practical to construct the imaginary potentials as there are analytical
expressions for the corresponding dispersion integrals~\cite%
{Vanderkam00,Quesada03}.
As the imaginary spin-orbit component is
generally needed only at high energies, we have kept the form employed in Ref.~\cite{Mueller11}
\begin{equation}
W^{so}(E)= A^{so}  \frac{(E-\varepsilon_F)^4}{(E-\varepsilon_F)^4+(B^{so})^4} .
\label{eq:ImSO}
\end{equation}%
Referring to Eq.~(\ref{eq:sdisprel}) all ingredients of the DOM have now been identified and their functional form described.
In addition to the HF contribution and the absorptive potentials we also include the dispersive real part from all imaginary contributions according to Eq.~(\ref{eq:sdisprel}).

The solution of the Dyson equation below the Fermi energy was introduced in Ref.~\cite{Dickhoff10} and more details can be found there.
The scattering wave functions are generated with the iterative procedure outlined in Ref.~\cite{Michel09} leading to a modest increase in computer time as compared to the use of purely local potentials.
Neutron and proton potentials are kept identical in the fit except for the Coulomb potential.
Included in the present fit are the same elastic scattering data and level information considered in Ref.~\cite{Mueller11}.
We refer to that paper for references to these data.
In addition, we now include the charge density of ${}^{40}$Ca as given in Ref.~\cite{deVries1987} by a sum of Gaussians in the determination of the DOM parameters. 
The calculation of the charge density requires a rescaling of the calculated density matrix from the $A-1$ to the $A$-body system as in Ref.~\cite{MuSi04}.
Data from the $(e,e'p)$ reaction at high missing energy and momentum obtained at Jefferson Lab for ${}^{12}$C~\cite{Rohe04}, ${}^{27}$Al, ${}^{56}$Fe, and ${}^{197}$Au~\cite{Rohe04A} were incorporated as well.
We note that the spectral function of high-momentum protons per proton number is essentially identical for ${}^{27}$Al and ${}^{56}$Fe thereby providing a sensible benchmark for their presence in ${}^{40}$Ca. 
We merely aimed for a reasonable representation of these cross sections since their interpretation requires further consideration of rescattering contributions~\cite{Barbieri06}.
A detailed representation of these data would require our potentials to exhibit a energy-dependent geometry which would increase the computational effort exponentially as discussed in Ref.~\cite{Dickhoff10}.
We did not include the results of the analysis of the $(e,e'p)$ reaction from NIKHEF~\cite{Kramer89} because the extracted  spectroscopic factors depend on the employed local optical potentials. We intend to reanalyze these data with our nonlocal potentials in a future study.

The numerical values of all parameters together with a reference to the appropriate equations are given in tables in the next section.

\section{Results}
\label{sec:Results}
\begin{figure}[tbp]
\includegraphics*[scale=0.6]{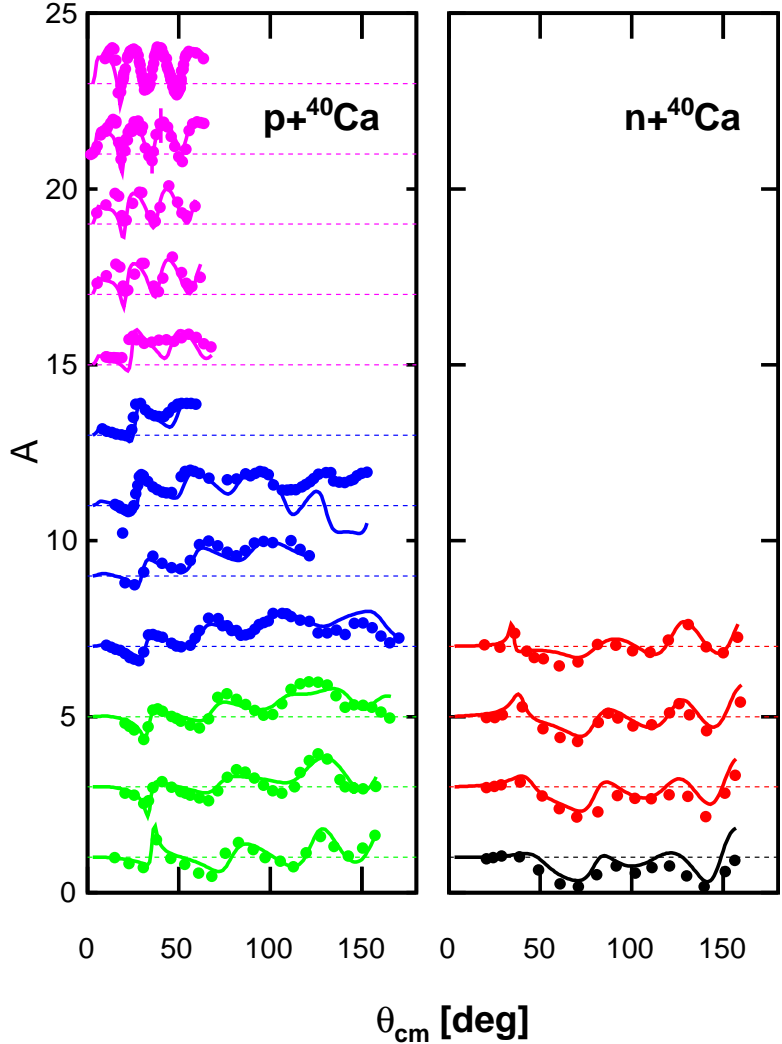}
\caption{Calculated and experimental analyzing powers. Panels shows results for \textit{n}+$%
^{40}$Ca and \textit{p}+ $^{40}$Ca. Data for each energy are offset for clarity with the lowest energy at the bottom and highest at the top of each frame. References to the data are given in Ref.~\protect\cite{Mueller11}.}
\label{fig:analyzing}
\end{figure}

All results are presented in Ref.~\cite{Hossein13b} except for the analyzing powers which are shown in Fig.~\ref{fig:analyzing}.
The quality of the fit is the same as in our previous fits with local potentials~\cite{Charity07,Mueller11}.

We note that we have continued to separate the geometry dependence of the potentials from their energy dependence to avoid a prohibitive increase in computational effort.
The constraint of the number of particles was incorporated to include contributions from $\ell = 0$ to 5.
Such a range of $\ell$-values generates a sensible convergence with $\ell$ when short-range correlations are included as in Ref.~\cite{Dussan11}. 
We obtain 19.88 protons from all $\ell = 0$ to 5 partial wave terms including $j = \ell \pm \frac{1}{2}$ and 19.79 for neutrons.
This is within the error we assigned to the particle number of 1\%.
If in future higher $\ell$-values are included, we expect a slight but not essential change in the fitted parameters.   

\begin{table}[tpb]
\caption{Fitted parameter values for the proton and neutron nonlocal HF potential in ${}^{40}$Ca. The table also contains the number of the equation that defines each individual parameter.}
\label{Tbl:fit}%
\begin{ruledtabular}
\begin{tabular}{ccc}
parameter &                                  value & Eq. \\
\hline
$V^0_{HF}$ [MeV]		& 100.06 & (\ref{eq:HFvol}) \\
$r^{HF}$ [fm]                      &  1.10   & (\ref{eq:HFvol}) \\
$a^{HF}$ [fm]                       &  0.68   & (\ref{eq:HFvol}) \\
$\beta_{vol_1}$ [fm]		& 0.66 & (\ref{eq:HFvol}) \\
$\beta_{vol_2}$ [fm]		& 1.56 & (\ref{eq:HFvol}) \\
$x_1$ 				& 0.48	& (\ref{eq:HFvol}) \\
$V^0_{wb}$ [MeV]                 &  15.0   & (\ref{eq:wb}) \\
$\rho_{wb}$ [fm]                 &  2.06   & (\ref{eq:wb}) \\
$\beta_{wb}$ [fm]                 &  1.10   & (\ref{eq:wb}) \\
\end{tabular}
\end{ruledtabular}
\end{table}
The HF parameters are shown in Table~\ref{Tbl:fit}. We note that the number of parameters is the same as for the local HF potential employed in Ref.~\cite{Mueller11}.
The appearance of two nonlocality parameters is also reasonable as the HF term calculated according to Eq.~(\ref{eq:disprel}) at the Fermi energy $\varepsilon_F$ receives dispersive contributions from energies that emphasize long-range (surface) as well as shorter-range (volume) terms.
The corresponding equations where these parameters are introduced have also been listed in this table.
The spin-orbit parameters are gathered in Table~\ref{Tbl:SO}.
We have kept the parameters of the imaginary component fixed to the values found in Ref.~\cite{Mueller11}  as indicated by the asterisk.

\begin{table}[tpb]
\caption{Fitted parameter values for the local HF and imaginary spin-orbit potentials in ${}^{40}$Ca. For those parameters indicated by an asterisk we kept the same values as in Ref.~\protect\cite{Mueller11}. The table also contains the number of the equation that defines each individual parameter.}
\label{Tbl:SO}%
\begin{ruledtabular}
\begin{tabular}{ccc}
parameter&                                  value & Eq. \\
\hline
$V^{so}_0[MeV]$                    & 6.03    & (\ref{eq:HFso})\\
$r^{so}$ [fm]                      &  1.02   & (\ref{eq:HFso}) \\
$a^{so}$ [fm]                       &  0.66   & (\ref{eq:HFso}) \\
$A^{so}$ [MeV]                     & -3.65(*) & (\ref{eq:ImSO}) \\
$B^{so}[MeV]$                      &  208(*) &  (\ref{eq:ImSO})     \\
\end{tabular}
\end{ruledtabular}
\end{table}

The parameters pertaining to volume absorption are displayed in Table~\ref{Tbl:ImVol}.
We note that volume absorption plays different roles in the fit.
At positive energies above 50 MeV it provides the absorption necessary to describe elastic, total, and reaction cross sections of protons and neutrons. 
In addition, it removes single-particle strength from the Fermi sea and moves it to those energies.
Conversely, and to make up for the loss of particles, the volume absorption below the Fermi energy moves high-momentum strength to energies below the location of the lowest $s_{1/2}$ orbit in accord with Jefferson Lab data and theoretical predictions for finite nuclei~\cite{Muther94,Dussan11}. 
Motivated by the theoretical work of Refs.~\cite{Waldecker2011,Dussan11}, we allow for different nonlocalities above and below the Fermi energy.
The values of the nonlocality parameters $\beta$ appear reasonable and range from 0.64 fm above to 0.81 fm below the Fermi energy.
These parameters together with integral properties of these absorbing potentials are critical in ensuring that particle number is adequately described. 
While we observe small differences for the radius and diffuseness parameters other parameters show very minor differences and could have been kept identical above and below the Fermi energy in the fit.
The only exception appears to be the parameters $E^{\pm}_a$ but they play a somewhat different role according to Eq.~(\ref{eq:Wnmnl}).
Noteworthy is the extended energy domain for volume absorption below $\varepsilon_F$ to accommodate the Jefferson Lab data as indicated by the parameter $E^-_a$.
It is important to generate correct values for particle number in order to obtain quality results for data below the Fermi energy.
This can only be achieved by including nonlocal absorptive potentials that are also constrained by the high-momentum spectral functions.
With local absorption we are not capable to either generate a particle number of 20 or describe the charge density accurately~\cite{Dickhoff10}, while simultaneously describing the positive-energy scattering data.
\begin{table}[tpb]
\caption{Fitted parameter values for proton and neutron potentials in ${}^{40}$Ca that determine volume absorption. 
The table also contains the number of the equation that defines each individual parameter.}
\label{Tbl:ImVol}%
\begin{ruledtabular}
\begin{tabular}{ccc}
parameter &                                  value & Eq. \\
\hline
$r_{vol}^+$ [fm] 		& 1.37 	& (\ref{eq:imnl}) \\ 
$a_{vol}^+$ [fm] 		& 0.68 	& (\ref{eq:imnl}) \\ 
$\beta_{vol}^+$ [fm]		& 0.64	& (\ref{eq:imnl}) \\ 
$r_{vol}^-$ [fm] 			& 1.44 	& (\ref{eq:imnl}) \\ 
$a_{vol}^-$ [fm] 		& 0.50 	& (\ref{eq:imnl}) \\ 
$\beta_{vol}^-$	[fm]		& 0.81	& (\ref{eq:imnl}) \\
$A^{vol}_+$ [MeV]		& 7.74	& (\ref{eq:volumeS}) \\
$B^{vol}_+$ [MeV]		& 25.87	& (\ref{eq:volumeS}) \\
$E^{vol}_{p+}$ [MeV]	& 13.59	& (\ref{eq:volumeS}) \\
$A^{vol}_-$ [MeV]		& 9.50	& (\ref{eq:volumeS}) \\
$B^{vol}_-$ [MeV]		& 27.29	& (\ref{eq:volumeS}) \\
$E^{vol}_{p-}$ [MeV]		& 5.50	& (\ref{eq:volumeS}) \\
$\alpha\ [\textrm{MeV}^{-1/2}]$      &  0.125   & (\ref{eq:Wnmnl}) \\
$E_a^+$ [MeV]			& 19.59	& (\ref{eq:Wnmnl}) \\
$E_a^-$ [MeV]			& 120	& (\ref{eq:Wnmnl}) \\
\end{tabular}
\end{ruledtabular}
\end{table}

\begin{table}[tpb]
\caption{Fitted parameter values for proton and neutron potentials in ${}^{40}$Ca that determine surface absorption. The table also contains the number of the equation that defines each individual parameter.}
\label{Tbl:ImSur}%
\begin{ruledtabular}
\begin{tabular}{ccc}
parameter &                                  value & Eq. \\
\hline
$r_{sur}^+$ [fm] 		& 1.15 	& (\ref{eq:imnl}) \\ 
$a_{sur}$ [fm] 		& 0.60(*) 	& (\ref{eq:imnl}) \\ 
$\beta_{sur}^+$ [fm]		& 0.94	& (\ref{eq:imnl}) \\ 
$r_{sur}^-$ [fm] 			& 1.19 	& (\ref{eq:imnl}) \\ 
$\beta_{sur}^-$	[fm]		& 2.07	& (\ref{eq:imnl}) \\
$A^{sur}_+ $ [MeV]                     &  12.31   & (\ref{eq:paranl}) \\
$B^{sur}_{+s1}$ [MeV]                     &  13.87   & (\ref{eq:paranl}) \\
$B^{sur}_{+s2}$ [MeV]                     &  36.62   & (\ref{eq:paranl}) \\
$C^{sur}_+$ [MeV]                        &  17.21   &  (\ref{eq:paranl}) \\
$A^{sur}_- $ [MeV]                     &  7.21   & (\ref{eq:paranl}) \\
$B^{sur}_{-s1}$ [MeV]                     &  14.34   & (\ref{eq:paranl}) \\
$B^{sur}_{-s2}$ [MeV]                     &  25.46   & (\ref{eq:paranl}) \\
$C^{sur}_-$ [MeV]                   &  17.33   &  (\ref{eq:paranl}) \\
\end{tabular}
\end{ruledtabular}
\end{table}

Surface absorption parameters are collected in Table~\ref{Tbl:ImSur}. 
Surface absorption is essential to describe low-energy scattering data and reflects the coupling of a nucleon to low-lying collective states and giant resonances~\cite{Dickhoff04}.
Expectations are that such a coupling is more or less symmetric above and below the Fermi energy~\cite{Mahaux91} and was therefore assumed in earlier DOM applications.
The \textit{ab initio} self-energy calculations of Ca isotopes reported in Ref.~\cite{Waldecker2011} and based on the Faddeev random phase approximation~\cite{Barbieri1,Barbieri2} indicate that this symmetry may not be perfectly adhered to.
We have therefore abandoned a strict symmetry assumption for the surface-absorption parameters but the values displayed in Table~\ref{Tbl:ImSur} indicate that most parameters acquire very similar values above and below the Fermi energy and a symmetric version may be restored in future work.
Since the emphasis of surface absorption is around the Fermi energy, it removes about as much single-particle strength from below to above the Fermi energy as the other way around and therefore plays a less important role in reproducing particle number.
It is however important below the Fermi energy to generate the appropriate widths of more deeply-bound levels and allow for the focussing of single-particle states near the Fermi energy due to its real dispersive correction~\cite{Mahaux91,Charity06}.

In conclusion, the work reported in Ref.~\cite{Hossein13b} has demonstrated that it is possible to generate an accurate description of all data that can be accessed by the single-particle propagator of a nucleus like ${}^{40}$Ca. 
One obvious extension of this work requires a fresh analysis of the $(e,e'p)$ data for this nucleus~\cite{Kramer89}.
We note that the standard analysis of this reaction employs local optical potentials and overlap functions also generated by a local Woods-Saxon well.
With such an analysis it may be possible to assess whether the spectroscopic factors for the valence levels generated by the nonlocal DOM will lead to a consistent description of the $(e,e'p)$ cross sections when all ingredients are provided by the DOM.
Transfer reactions have been successfully analyzed with local DOM ingredients in Ref.~\cite{Nguyen2011}.
Inclusion of nonlocal potentials may lead to further insights and new results including those for exotic nuclei. 
We therefore plan to extend the DOM analysis to nuclei with $N \ne Z$ in particular ${}^{48}$Ca and ${}^{208}$Pb for which accurate charge densities are available~\cite{deVries1987}.
The possibility to explore the empirical contribution of three-body forces to the binding energy of nuclei as made possible by our current analysis, also remains an intriguing project~\cite{Hossein13b}.

This work was supported by the U.S. Department of Energy,
Division of Nuclear Physics under grant DE-FG02-87ER-40316 and the U.S.
National Science Foundation under grants PHY-0968941 and PHY-1304242.

\bibliography{DOMbib_W}

\end{document}